\documentclass[twocolumn,showpacs,preprintnumbers,amsmath,amssymb,
	superscriptaddress]{revtex4}

\usepackage{graphicx}
\usepackage{dcolumn}
\usepackage{bm}

\begin{document}

\preprint{}

\title{Study of the ground state properties of 
	$\mathrm{LiHo_xY_{1-x}F_4}$
	using $\mu$SR}

\author{J. Rodriguez}
 \affiliation{Department of Physics and Astronomy,
   McMaster University, 1280 Main St. W.,
   Hamilton, ON, Canada, L8S 4M1} 
 \email {rodrigja@mcmaster.ca}

\author{A.A. Aczel}
 \affiliation{Department of Physics and Astronomy,
   McMaster University, 1280 Main St. W.,
   Hamilton, ON, Canada, L8S 4M1} 

\author{J.P. Carlo}
 \affiliation{Department of Physics,
   Columbia University, 
   538 W. 120th St., New York, NY, 10027}

\author{S.R. Dunsiger}
 \affiliation{Department of Physics and Astronomy,
   McMaster University, 1280 Main St. W.,
   Hamilton, ON, Canada, L8S 4M1} 
 \affiliation{Physics Department,
	James Franck Strasse 1, Munich Technical University,
	D-85748 Garching, Germany.}

\author{G.J. MacDougall}
 \affiliation{Department of Physics and Astronomy,
   McMaster University, 1280 Main St. W.,
   Hamilton, ON, Canada, L8S 4M1} 
 \affiliation{Oak Ridge National Laboratory, 
 	Oak Ridge, Tennessee 37831, USA.}

\author{P.L. Russo}
 \affiliation{Department of Physics,
   Columbia University, 
   538 W. 120th St., New York, NY, 10027}

\author{A.T. Savici}
 \affiliation{Department of Physics,
   Columbia University, 
   538 W. 120th St., New York, NY, 10027}
 \affiliation{Oak Ridge National Laboratory, 
 	Oak Ridge, Tennessee 37831, USA.}

\author{Y.J. Uemura}
 \affiliation{Department of Physics,
   Columbia University, 
   538 W. 120th St., New York, NY, 10027}

\author{C.R. Wiebe}
 \affiliation{Physics Department,
   Florida State University,
   315 Keen Building, Tallahassee, FL 32306-4350}

\author{G.M. Luke}
 \affiliation{Department of Physics and Astronomy,
   McMaster University, 1280 Main St. W.,
   Hamilton, ON, Canada, L8S 4M1}

\date{\today}

\begin{abstract}
$\mathrm{LiHo_xY_{1-x}F_4}$ is an insulating system
where the magnetic Ho$^{\rm 3+}$ ions have an Ising character, and
interact mainly through magnetic dipolar fields. 
We used the muon spin relaxation technique to study the
nature of the ground state for samples with x=0.25, 0.12,
0.08, 0.045 and 0.018. In contrast with some previous works, we have
not found any signature of canonical spin glass behavior 
down to $\approx$15mK. Instead, below $\approx$300mK
we observed dynamic magnetism characterized by a single correlation time
with a temperature independent 
fluctuation rate. We observed that this low temperature 
fluctuation rate increases with $x$ up to 0.08, above which
it levels off. The 300mK energy scale corresponds to the Ho$^{3+}$ hyperfine
interaction strength, suggesting that the hyperfine interaction may be intimately 
involved with the spin dynamics in this system. 
\end{abstract}


\maketitle

Ising models play a central role in our
understanding of magnetic systems and their phase transitions. 
Their importance stems from their simplicity with respect
to other models, and in the fact that they reproduce many 
observed physical phenomena (e.g. glassiness and
quantum phase transitions). For T$\lesssim$2K and x$<$1,
$\mathrm{LiHo_xY_{1-x}F_4}$ is thought to be a physical 
realization of the random transverse field Ising
model with dipolar-magnetic interactions (plus a smaller
nearest neighbor antiferromagnetic exchange interaction)
\cite{schechter2005,tabei2006}.
For 0.25$\lesssim$x$\leq$1 the ground state of the
system is a ferromagnet 
\cite{cooke1975, battison1975, reich1990};
and for x$\lesssim$0.25 
enough randomness is introduced in the system such that 
long range ferromagnetic order is destroyed 
\cite{reich1990,jonsson2007,anconatorres2008}. 
It is natural to expect that in this last diluted 
regime, the long ranged 
dipolar interaction (which can be antiferromagnetic
for many bonds) together with the quenched chemical
disorder produce a spin glass ground state.
Surprisingly, the nature
of the ground state for x$\lesssim$0.25 has 
been the topic of a heated  debate at both the 
experimental and theoretical levels. 

Experimentally it was found that for 
0.1$\lesssim$x$\lesssim$0.25 the non-linear
AC susceptibility ($\chi_3$) peaks as a function 
of temperature, and that this peak gets rounded
upon the application of an external magnetic field
perpendicular to the Ising axis 
\cite{wu1993,anconatorres2008}. 
These measurements were interpreted by the authors as a 
transition to a low temperature spin glass state. 
This interpretation was supported by 
a numerical calculation \cite{tam2009}, and 
the rounding of the peak in the presence of an external
field was proposed to be a consequence of 
field induced random fields 
\cite{schechter2005,tabei2006}. Other 
researchers though, pointed out that a critical
analysis of $\chi_3$ using only data above the peak 
(in equilibrium) indicates that there is no transition
into a spin glass state at any finite temperature and 
transverse field
\cite{mattsson1995,jonsson2007,jonsson2008},
in agreement with previous numerical calculations
\cite{snider2005,biltmo2007,biltmo2008}.

At a lower doping (x=0.045), one research group 
observed that 
the frequency response of the linear AC susceptibility 
is narrower than that of 
a spin glass, and this
was interpreted as a splitting of the system into 
clusters of spins which behave as single harmonic
oscillators \cite{ghosh2002}.
In contrast, this narrowing was not
observed in the measurements from another group which
used a different sample with the same doping level.
Instead, the temperature dependence of this $\chi_{\rm AC}$ data 
was shown to be compatible with that of
a spin glass with a transition temperature 
lower than that achieved by the measurements
\cite{quilliam2008}. 

At this time there is no consensus 
on the nature of the ground state for 
x$\lesssim$0.25. One of the main reasons for this is
the lack
of experimental data with different probes, as available 
data is mostly on magnetic susceptibility and specific
heat. With this in mind, we report in this letter
muon spin relaxation ($\mu$SR) \cite{summerschool} 
measurements in a series of samples which span
the whole diluted regime (x=0.25, 0.12, 0.08, 0.045 and
0.018). Our measurements do not present any of the signatures
of canonical spin glass freezing; and an analysis of our data
using a Kubo-Toyabe model \cite{uemura1999} reveals that 
the Ho$\rm ^{3+}$ ions slow down 
with decreasing temperature down to $\approx$300mK. Below 
this point and down to $\approx$15mK, the fluctuation
rate of the Ho$\rm ^{3+}$ ions is observed to be temperature 
independent \cite{rodriguez2006}. Our analysis also shows that the
low temperature fluctuation rate of the Ho$\rm ^{3+}$ magnetic
moments increases with x up to 0.08 after which it levels off. 

Our samples are single crystals purchased from 
TYDEX J.S.Co. (St. Petersburg). Pieces from the main
crystals were placed in the sample holder in such a way
that the externally applied magnetic field was
perpendicular to the Ising axis. The $\mu$SR measurements
were performed at the M15 and M20 beam lines of 
TRIUMF (Canada) in the Longitudinal Field (LF) configuration.
In this configuration the initial muon spin direction is
along the external magnetic field and therefore perpendicular
to the Ising axis. For the measurements at M15, the samples 
were mounted on a silver sample holder using ``Apiezon N" 
grease for thermal contact. This holder was then screwed 
to the mixing chamber of a dilution refrigerator. In this 
device the temperature of the samples
was typically varied between 15mK and 3K, while the external
field was scanned up to 0.2T. In the M20 beam line the 
temperature was controlled with a helium 
flow cryostat in the range from 1.8K to 100K, and the samples 
were mounted using a low background sample holder.

Upon cooling from T$\approx$20K, the relaxation of the 
signal increases due to slowing down of the 
magnetic Ho$^{\rm +3}$ ions into the $\mu$SR time 
window \cite{rodriguez2006,graf2007}.
The increase of the relaxation upon cooling is 
monotonic down to base temperature.
Since a spin glass ground state is expected to be observed 
at these dilution levels,
we analyzed our low LF data using a 
power-exponential fitting function: $\exp(-(\lambda t)^\beta)$. 
This function 
has been successfully used to study $\mu$SR lineshapes of 
disordered spin systems, including
spin glasses above ${\rm T_g}$ \cite{campbell1999}. 

\begin{figure}[tbp]
\includegraphics[angle=0,width=8.6cm]{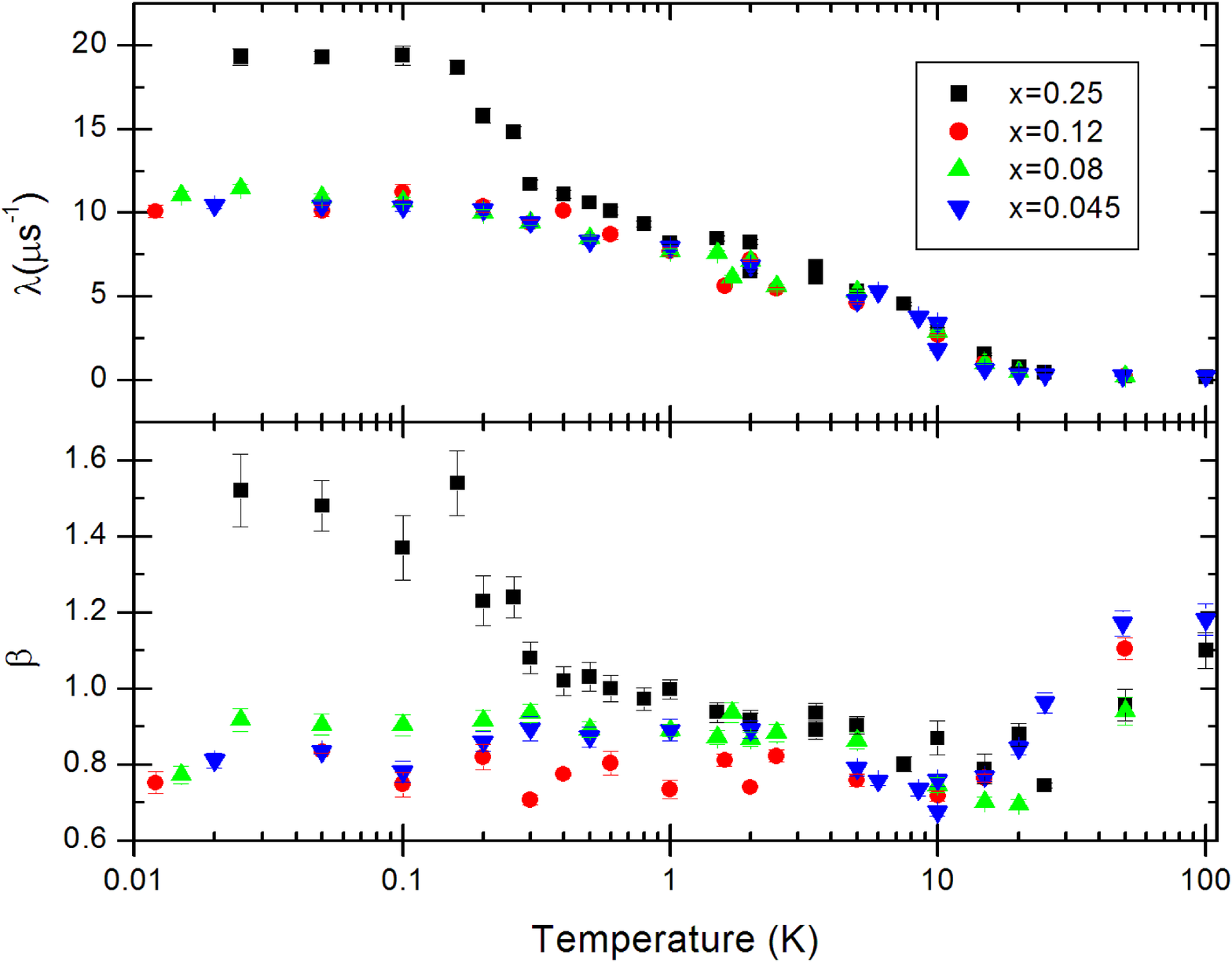}
\caption{\label{f2} Analysis of the low LF data using 
	a power-exponential fitting function.
	The top panel shows the relaxation rate
	of the signal $\lambda$, and the lower one 
	the exponent $\beta$. Color online.}
\end{figure}

The fit values for $\lambda$ are shown in Figure \ref{f2}.
We will argue later that upon cooling, the increase in $\lambda$ 
around 200mK
(most noticeable in the x=0.25 data) is produced by a further
slowing down of the magnetic moments of the system. 
This figure also shows the fit values for the power
$\beta$. It can be seen that upon cooling, this parameter
has a minimum at around $\approx$10K, which is associated
with the slowing down of the very fast fluctuating Ho moments
into the $\mu$SR time window \cite{graf2007}. Below this 
temperature, $\beta$ grows monotonically and it stabilizes 
at $\approx$0.85 for the x=0.12, 0.08 and 0.045 systems, and at
$\approx$1.5 for x=0.25. If a glassy behavior was to be observed, 
this parameter should monotonically decrease upon cooling and reach
a minimum of 1/3 just above the freezing temperature 
$\mathrm{T_g}$ \cite{campbell1999}. 
An upper estimate for $\mathrm{T_g}$ in 
$\mathrm{LiHo_xY_{1-x}F_4}$ can be obtained using the mean field
expression: $\mathrm{T_{mf} \approx T_g \approx x T_c}$ 
($\mathrm{T_c}$ is
the critical temperature of the ferromagnetic x=1 system which is 
1.54K). These
estimated temperatures are shown in Table \ref{t1}, together with
the temperature at which the magnetic specific heat peaks 
\cite{quilliam2007}. Figure \ref{f2} shows
that none of our samples presents a minimum of $\beta$ at around
${\rm T_{mf}}$ or ${\rm T_{peak}}$. This fact together
with the observation that no change in the shape of the signals is
observed upon further cooling from $\approx$100mK, indicates that
these systems do not have a canonical spin glass ground state.

\begin{table}[tbp]
\begin{center}
\begin{tabular}{|c|c|c|c|c|} 
\hline
x & ${\rm T_{mf}}$ (mK) & ${\rm T_{peak}}$ (mK)
& $a$ or $\Delta$ (${\rm \mu s^{-1}}$) & 
$\nu_0$ (${\rm \mu s^{-1}}$) \\
\hline
0.25 & 390 & - & 17.7(2) & 15.6(8) \\
\hline
0.12 & 180 & - & 11.8(4) & 20(1) \\
\hline
0.08 & 120 & 120 & 12.6(1) & 20(2) \\
\hline
0.045 & 60 & 130 & 9.6(1) & 10.5(1) \\
\hline
0.018 & 30 & 110 & 4.5(2) & 0.73(2) \\
\hline
\end{tabular}
\caption[]{Estimated freezing temperature (${\rm T_{mf}}$),
	position of the peak of the specific heat (${\rm T_{peak}}$) 
	\cite{quilliam2007},
	characteristic strength of the internal field ($a$ or $\Delta$),
	and low temperature fluctuation rate of Ho$^{\rm 3+}$ ions 
	($\nu_0$), for each of the studied samples \label{t1}}
\end{center}
\end{table}

The low temperature data from the x=0.018 system could not be 
properly fit by a power-exponential since
the $\mu$SR signal develops a shoulder at low temperatures. 
This type of behavior indicates that fluctuations
of the magnetic moments in the x=0.018 system are slow.
We decided to further analyze our data using Dynamical Kubo-Toyabe 
(DKT) polarization functions \cite{uemura1999}. The value of 
$\beta$ for the x=0.25
system at low temperature is approximately 1.5 (Figure \ref{f2})
which indicates that, magnetically speaking, the system rather dense.
Then,we used Gaussian DKT functions to fit this data. On the 
other hand, the low temperature values of $\beta$ for the other 
systems are $\approx$0.85, indicating that
these systems are magnetically diluted.
Then, we used Lorentzian DKT functions to fit the the data from
the x=0.12, 0.08, 0.045 and 0.018 systems. 
The fitting parameters of the Lorentzian (Gaussian) model are 
$a$ ($\Delta$) and $\nu$. $a$ and $\Delta$ represent the 
characteristic size of the internal magnetic field at the 
muon site, while $\nu$ is the fluctuation rate of the internal
field (or the inverse correlation time of the local magnetic
field, that is: $\langle B(0)B(t) \rangle \propto \exp (-\nu t)$).
We should mention that we attempted to analyze our data using other
microscopic models (such as the the spin glass 
function in Reference \cite{uemura1985}) but none
them produced sensible (or physical) results \cite{rodriguez2009}. 

The fit of the low LF data using the DKT functions were performed 
by fixing $a$ ($\Delta$ for the x=0.25 system) to the
value found at base temperature, and then letting only the 
fluctuation rate $\nu$ vary as a function of temperature. The fits
with the DKT produced sensible results at the qualitative level
\cite{rodriguez2009}. The values of $a$ ($\Delta$ for the x=0.25
system)
are shown in Table \ref{t1}. These values were found to 
roughly follow the $\sqrt{{\rm x}}$ trend expected for low values of x
\cite{rodriguez2009}.

\begin{figure}[tbp]
\includegraphics[angle=0,width=8.6cm]{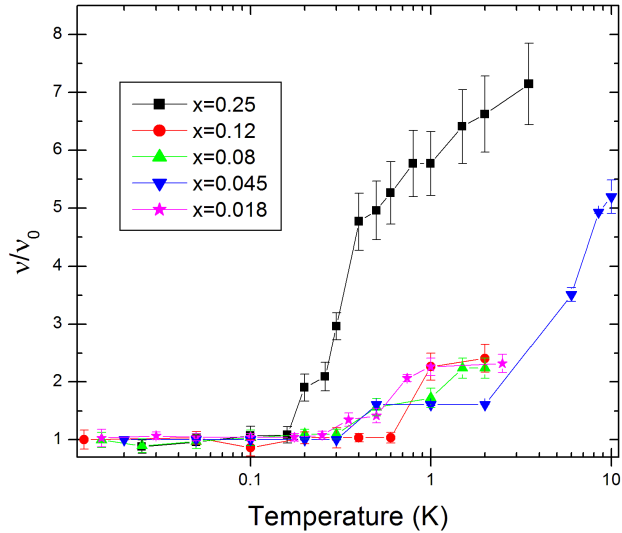}
\caption{\label{f4} Fitted values of $\nu$ divided by its
	low temperature average value ($\nu_0$) as a 
	function of temperature and doping level. Color online.}
\end{figure}

Figure \ref{f4} shows the value of $\nu/\nu_0$ as a function
of temperature for all our samples, where $\nu_0$ is
the average value of $\nu$ in the low temperature range where
it is constant. It can be seen that, as the 
temperature is lowered the fluctuation rate of the ions decreases;
and below a temperature $\mathrm{T^*}$ this fluctuation rate
becomes temperature independent. $\mathrm{T^*}$ does not seem to 
have a clear dependence on x, and has a typical
value of 300mK. We note that this temperature is approximately the 
same size as the hyperfine interaction energy scale of the 
Ho$^{\rm 3+}$ ions ($\approx$200mK) \cite{giraud2001}.

The fit values of $\nu_0$ are shown in Table \ref{t1}.
It can be seen that $\nu_0$ increases with Ho concentration 
until approximately 0.08, above which it levels off. It is
interesting to notice that the point of the x-T phase diagram
where the characteristic strength of the dipolar interaction 
is smaller than that of the hyperfine interaction \cite{comment1}
is 0.13, which is close to 0.08, the point where we observed
$\nu_0$ to flatten.

\begin{figure}
\includegraphics[angle=0,width=8.6cm]{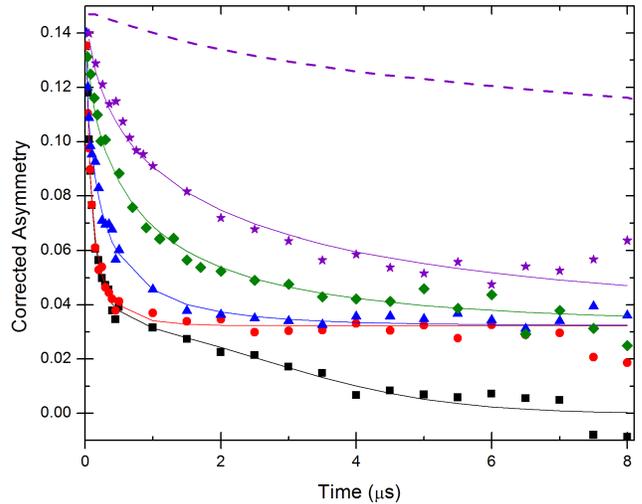}
\caption{\label{f1} Corrected asymmetry for
	$\mathrm{LiHo_{0.12}Y_{0.88}F_4}$ at 12mK and 
	different LF. Zero field - black squares, 0.01T - red
	circles, 0.05T - blue triangles, 0.1T - green
	vertical squares, and 0.2T - violet stars. The
	continuous lines are fits to Lorentzian DKT 
	functions ($a$=${\rm 12 \mu s^{-1}}$, 
	$\nu$=${\rm 20 \mu s^{-1}}$). The partial 
	decoupling for LF $\ge$ 0.01T is due to a 
	temperature-independent background present in
	the zero field data. The violet dashed line is the
	expected DKT line shape in 0.2T if the system was
	quasi-static ($a$=${\rm 12 \mu s^{-1}}$ and 
	$\nu$=${\rm 1.2 \mu s^{-1}}$). Color online.}
\end{figure}

The suitability of the DKT model to describe the low temperature
behavior of $\mathrm{LiHo_xY_{1-x}F_4}$ was tested with the high
LF data (see Figure \ref{f1}). Using the low LF fit 
parameters, we only increased
the field in the DKT model and the resulting functions were observed
to follow satisfactorily the experimental data up to 
$\approx$0.2T for the x=0.25, 0.12, 0.08 and 0.045 samples.
The LF scan in 
Figure \ref{f1} clearly shows that the system is dynamic at low
temperature. If the magnetic environment was static (frozen),
it is expected that 
a LF of 0.2T would decouple the $\mu$SR signal by about 85\%
(corresponding to an asymmetry of 0.14 in the figure).
Instead, at this field the signal 
relaxes much below this point, evidencing the dynamic environment. 
Further more, the dashed line in the figure shows that the
system is not even quasi-static. The fact that
the $\mu$SR signals for LF$<$0.2T are satisfactorily described by 
the DKT model with a fixed value of $\nu_0$, 
indicate that the external field does not have a big 
effect on this parameter (as expected since the 
Ising levels are not coupled for LF$\ll$2T 
\cite{schechter2005}).

The experimental lineshape at intermediate LF
($\approx$200G) in the x=0.018 system was not followed as closely
as that for
the denser systems.  We believe that this is due to the
internal field distribution of the system being stretched along
the Ising axis \cite{rodriguez2009}, instead of the spherical one
that the DKT function assumes. This deformation
has a big effect on the way the static signal is decoupled, and
therefore its effects are most appreciable at this doping due to
its low value of $\nu_0$ (compared to the parameter
$a$, see Table \ref{t1}). 

It should be remarked that our data shows the
same qualitative behavior for all the samples over
the explored temperature-field space. This is in 
contrast with the claim that the
x=0.045 system has a physically different ground state
from the x=0.167 one \cite{reich1990}. 
We should note though that specific heat and AC susceptibility 
measurements of our x=0.045 sample \cite{quilliam2007,quilliam2008} 
are different from those reported in References
\cite{reich1990,ghosh2002,ghosh2003}, opening the door to the
possibility that the the difference lies at the sample level. 
More measurements with other samples is needed to clarify 
this point. 

It is interesting to compare our results with those 
from other experimental techniques. Non-linear 
$\chi_{\rm AC}$ measurements in the x=0.198 and 0.167
systems show a peak at $\approx$140mK and 
$\approx$130mK respectively \cite{anconatorres2008}. 
These temperatures, as well as those at which the magnetic
specific heat peaks (Table \ref{t1}), are in the range where
we observe the onset of temperature independent
fluctuations, between 100mK and 600mK.
Then, it is possible that the spin glass behavior of 
$\mathrm{LiHo_xY_{1-x}F_4}$ is associated with the
slowing of the magnetic moments down to $\approx$300mK; 
but note that the temperature independent fluctuations
at low temperature indicate that the freezing is not completed
(Figure \ref{f4}). This is in agreement with the
absence of a spin glass transition obtained from a critical
analysis of other non-linear $\chi_{\rm AC}$ measurements 
\cite{jonsson2007,jonsson2008}.

It is natural to expect a spin glass ground state 
in $\mathrm{LiHo_xY_{1-x}F_4}$, since it posses the required
characteristics:
frustration, introduced by the dipolar interactions, and quenched
disorder from the random dilution. As shown before, the ground 
state of the system is dynamic, and therefore is not a classical
spin glass. In this sense,
the observation of a dynamic ground state in 
$\mathrm{LiHo_xY_{1-x}F_4}$ is as
surprising as the observation of spin glassiness in the pyrochlore
$\mathrm{Tb_2Mo_2O_7}$ \cite{gaulin1992,dunsiger1996} 
(a similar compound, $\mathrm{Y_2Mo_2O_7}$, is also a spin
glass \cite{gingras1997,dunsiger1996}, but there is recent evidence 
for atomic position disorder in it \cite{greedan2009,booth2000}).
As the diluted dipolar Ising model does 
seem to have a spin glass ground state \cite{tam2009}, it is 
tempting to believe that the hyperfine interaction can be responsible
for the low temperature properties. The
importance of this term for $\mathrm{LiHo_xY_{1-x}F_4}$
has already been pointed out in References 
\cite{schechter2005, schechter2008_2}; and
we found that its energy scale coincides with the onset
of the temperature-independent dynamic ground state. Also, as
mentioned before, the hyperfine energy scale might also appear in
the dependence of $\nu_0$ with dilution. Nevertheless, it is not clear 
 how can the hyperfine term be responsible for
the observed dynamical behavior, since this term has the effect of 
preventing fluctuations between the Ising levels instead of 
promoting them \cite{schechter2005}.

In summary, our $\mu$SR measurements show that the 
ground state of $\mathrm{LiHo_xY_{1-x}F_4}$ for 
x$\le$0.25 is not that of a canonical spin glass. 
We have observed that the low temperature state of the
system is dynamical, and can be described satisfactorily 
by the stochastic
Kubo-Toyabe model (which assumes a single correlation
timescale). Using this model we have determined that the
fluctuation
rate of the Ho magnetic moments decreases as the temperature
is lowered until $\approx$300mK. Below this temperature
the fluctuation rate is temperature independent down to 
$\approx$13mK. The $\mu$SR signals
from all our samples exhibit the same qualitative behavior
as a function of temperature and doping which stands in contrast to the 
observation of an additional
``anti-glass" phase inferred from some $\chi_{\rm AC}$ 
measurements. The hyperfine energy scale apparently manifests itself
 in the measured dynamical properties, suggesting 
that the hyperfine interaction is involved in the 
interesting dynamic behavior of this system.

\begin{acknowledgments}
We would like to thank M.J.P. Gingras and M. Schechter
for having very interesting discussions with us and for 
their useful comments. Research at McMaster is supported
by NSERC and CIFAR.  We acknowledge financial support from NSF DMR-
05-02706 and DMR-08-06846 (Materials World Network program)
at Columbia. 
\end{acknowledgments}

\bibliography{lihoyf}

\begin{thebibliography}{33}
\bibitem[{com()}]{comment1}
\bibinfo{note}{The characteristic strength of the dipolar interaction becomes
  smaller than the hyperfine energy scale for: (1.54K)x$<$0.2K. Here 1.54K is
  the ferromagnetic transition temperature for the x=1 system, and 0.2K is the
  energy scale of the hyperfine interaction}.
\end{thebibliography}

\end{document}